\documentclass[prd,aps,twocolumn,showpacs,preprintnumbers,amsmath,amssymb]{revtex4}
\usepackage[dvips]{graphicx}
\usepackage{epsf}
\usepackage{amsmath}
\usepackage{amssymb}

\usepackage{graphicx}% Include figure files
\usepackage{dcolumn}% Align table columns on decimal point
\usepackage{bm}% bold math
\def\be{\begin{equation}}
\def\ee{\end{equation}}
\def\bea{\begin{eqnarray}}
\def\eea{\end{eqnarray}}

\begin{document}

\title{{Thermal Fluctuations and Bouncing Cosmologies}}

%\vspace{5mm}

\author{ Yi-Fu Cai$^1$\footnote{caiyf@ihep.ac.cn},
Wei Xue$^{2}$\footnote{xuewei@physics.mcgill.ca},
Robert Brandenberger$^{2,1,3,4}$\footnote{rhb@physics.mcgill.ca},
Xinmin Zhang$^{1,3}$\footnote{xmzhang@ihep.ac.cn}}

\affiliation{1) Institute of High Energy Physics, Chinese Academy of
 Sciences, Beijing 100049, P. R. China}

\affiliation{2) Department of Physics, McGill University, Montr\'eal, QC, H3A 2T8, Canada}

\affiliation{3)Theoretical Physics Center for Science Facilities
 (TPCSF), CAS, P. R. China }

\affiliation{4) Kavli Institute for Theoretical Physics, CAS, Beijing 100190, P. R. China}


\begin{abstract}
We study the conditions under which thermal fluctuations
generated in the contracting phase of a non-singular bouncing
cosmology can lead to a scale-invariant spectrum of
cosmological fluctuations at late times in the expanding
phase. We consider point particle gases, holographic gases and
string gases. In the models thus identified, we also study the
thermal non-Gaussianities of the resulting distribution of
inhomogeneities. For regular point particle radiation, we find
that the background must have an equation of state $w = 7/3$
in order to obtain a scale-invariant spectrum, and that the
non-Gaussianities are suppressed on scales larger than
the thermal wavelength. For Gibbons-Hawking
radiation, we find that a matter-dominated background yields
scale-invariance, and that the non-Gaussianities are large.
String gases are also briefly considered.

\end{abstract}

\maketitle

\section{Introduction}

In recent years, there has been a lot of interest in non-singular
bouncing cosmologies (see e.g \cite{Novello} for a recent
review with an extensive list of references). Such cosmologies
may be desirable since they resolve the singularity problem
of  the inflationary scenario, the current paradigm
of early universe cosmology. If inflation is realized by
making use of the potential energy of a scalar matter field
while treating space-tiime dynamics using the Einstein action,
then an initial cosmological singularity is unavoidable \cite{Borde}.

While it is possible that that in the context of a non-singular
bouncing cosmology a period of inflationary expansion \cite{Guth} is
realized after the bounce (see e.g. \cite{Cai} for an explicit model)
and the cosmological perturbations
observed today are generated as quantum vacuum fluctuations
in the inflationary phase \cite{ChibMukh}, it is also possible
to obtain fluctuations without requiring an inflationary phase
after the bounce. The reason is that all scales observed today
were at one point inside the Hubble radius at sufficiently early
times during the contracting phase. This is illustrated in
Figure 1, which shows the space-time plot in a non-singular
bouncing cosmology \footnote{Another possibility for obtaining a
scale-invariant spectrum of cosmological perturbations is in the context
of string gas cosmology \cite{BV}, in which it can be shown \cite{NBV} that
thermal string fluctuations in a quasi-static Hagedorn phase in the
early universe lead to a scale-invariant spectrum of both
scalar and tensor metric fluctuations, with a blue tilt of the tensor
modes as a characteristic signature \cite{BNPV1}
(see \cite{BNPV2,RHBSGrev} for recent reviews).}.

In the context of studies of bouncing cosmologies it has been
realized that fluctuations which are generated as quantum vacuum
perturbations and exit the Hubble radius during a matter-dominated
contracting phase lead to a scale-invariant spectrum of cosmological
fluctuations today \cite{FB2,Wands,Wands2} (see also \cite{Starob}),
and thus yield an alternative to cosmological inflation for explaining
the current observational data.

However, in a bouncing cosmology it is not manifest that perturbations
arise as quantum vacuum fluctuations. In inflationary cosmology
it can be argued that the exponential expansion of space during
the inflationary phase red-shifts all classical matter initially
present and leaves behind a matter vacuum. However, if the universe
starts out large and cold in a contracting phase, there does not
seem to be a reason to single out vacuum over thermal initial
conditions for the fluctuations.

Thermal fluctuations as the origin of structure in the universe were
considered in the context of a forever expanding cosmology, but
it was concluded that it was not possible to obtain a scale-invariant
spectrum of cosmological perturbations \cite{Joao}. However, as
realized in \cite{NBV}, thermal fluctuations of strings can give rise
to a scale-invariant spectrum if we abandon the assumption that
the universe is forever expanding. Specifically, in \cite{NBV} it
was shown that thermal string fluctuations in a quasi-static early
string phase yield a scale-invariant spectrum. Hence, there are
good reasons to expect that in a bouncing cosmology it might
be possible for thermal fluctuations to generate a spectrum whose
shape is in good agreement with observations.

In this paper we will study the conditions under which thermal
initial conditions for perturbations can lead to a scale-invariant
spectrum of cosmological fluctuations after the bounce \footnote{We
are assuming that the bounce is non-singular, as can be realized
if the gravitational action has the specific higher derivative form given in
\cite{Biswas} which is free of ghosts about Minkowski space-time
and can be shown to lead to non-singular bounces given various
forms of matter Lagrangians \cite{Biswas,Biswas2}. We also
assume that the bounce phase is short on the time scales of the
fluctuations of interest. This assumption will be used when
following the fluctuations through the bounce point.}. We find
various possibilities, depending on what kind of thermal
fluctuations we consider

If the fluctuations are generated by normal
particle radiation (with the usual equation of state $w_r = 1/3$,
where $w$ is the ratio $w = p / \rho$ of pressure $p$ to
energy density $\rho$), then the required equation of state of
the background is $w = 7/3$. For radiation in holographic
cosmology \cite{FS},
we find that a background equation of state of
$w = 0$  will lead to a scale-invariant
spectrum. String gases require a quasi-static early phase.

As realized \cite{Cai2009} in the study of perturbations
in the ``matter bounce", in which vacuum fluctuations exit
the Hubble radius during a matter-dominated phase of
contraction, inhomogeneities generated during a
contracting phase leave a distinctive imprint on the magnitude and
shape of the non-Gaussianities manifest in the three point function.
We will study these non-Gaussianities for thermal initial
conditions.

The outline of this paper is as follows: In the following section we study
the evolution of fluctuations in a bouncing universe. In Section 3, we
consider thermal initial conditions for the perturbations and ask
under which conditions a scale-invariant spectrum after the bounce
results. In Section 4 we then estimate the non-Gaussianities in the
resulting models.

\section{Fluctuations in a Bouncing Universe}

\begin{figure}[htbp]
\includegraphics[scale=0.3]{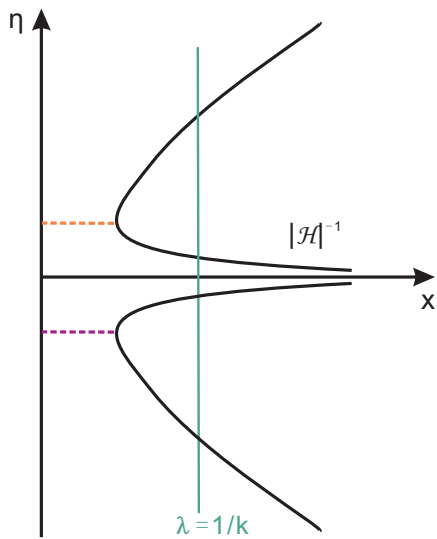}
\caption{A sketch of the evolution of scales in a bouncing
universe. The horizontal axis is the physical spatial coordinate,
the vertical axis is time. Plotted are the comoving Hubble radius
$|{\cal H}|^{-1}$ and the wavelength $\lambda$ of a fluctuation
mode with fixed comoving wavenumber $k$. } \label{Fig:sketch}
\end{figure}

It is useful to take a first glance at the route of cosmological
perturbations in a bouncing cosmology. As is depicted in
Figure \ref{Fig:sketch}, the wavelength of fluctuations becomes larger than the
Hubble radius in the contracting phase, and reenters in the expanding
phase. All that is required for this space-time sketch to apply is
that the equation of state parameter $w$ of the
background universe be larger than $-1/3$ at all times
except possibly around the bounce point. This requirement is
automatically satisfied for most forms of matter (the exception being
scalar field models which lead to inflation).
In Figure \ref{Fig:sketch} we plot the evolution of the physical length
corresponding to a fixed comoving scale. This scale is the
wavelength of the fluctuation mode $k$ which we will follow later.

In this paper we focus on adiabatic fluctuations and consider
matter without anisotropic stress. In this case, the
linearized fluctuations about a Friedmann-Robertson-Walker
background metric in longitudinal gauge
can be expressed as (see e.g. \cite{MFB} for a review of the
theory of cosmological perturbations)
\be
ds^2 \, = \, a(\eta)^2[(1+2\Phi)d\eta^2-(1-2\Phi)d\vec{x}^2]~,
\ee
where $\Phi(\vec{x}, \eta)$ is the generalized Newtonian gravitational potential
which characterizes the metric fluctuations, $\eta$ is conformal time and
$a(\eta)$ is the background scale factor \footnote{We are only considering
scalar metric fluctuations. However, it is important to point out that in cosmologies
with a contracting phase, vector \cite{BB} and tensor \cite{Wands2,LW}
fluctuations can also give a large contribution to the total angular power spectrum
of cosmic microwave background anisotropies.}.

In Fourier space,
the fluctuation variable $\Phi$ satisfies the following second order differential
equation
\begin{eqnarray}\label{perteq}
\Phi_k'' + 2\sigma {\cal H}\Phi_k' +
(c_s^2 k^2 - 2\epsilon{\cal H}^2 + 2\sigma{\cal H}^2)\Phi_k  \, = \, 0~,
\end{eqnarray}
where ${\cal H}\equiv a'/a$ is the comoving Hubble parameter, and
the prime denotes the derivative with respective to
$\eta$. The parameter $c_s$ is the sound speed, which we
take to be a free parameter.
Moreover, we have defined two useful parameters
\bea
\epsilon \, &\equiv& \, - \frac{\dot H}{H^2} \,\,\, {\rm and}  \\
\sigma \, &\equiv& \, -\frac{\ddot H}{2H\dot H}
\eea
which characterize the background
evolution. For a constant equation of state $w$, we have
\be
\epsilon \, = \, \sigma \, = \, \frac{3}{2}(1+w)
\ee
and thus the perturbation equation can be greatly simplified.

We will consider an equation of state $w \neq -1$ for which the
background scale factor evolves as a power of time. If $w > - 1/3$,
an equation of state which does not give accelerated expansion,
we can choose our origin of the time axis such that $t = 0$
corresponds to the bounce point. More specifically, we shall take
$t = 0$ to be the value of $t$ for which the Big Crunch singularity
would occur in the absence of the terms in the action which lead
to a non-singular bounce. In this case, we can also choose the
origin of the conformal time coordinate such that $t = 0$ corresponds
to $\eta = 0$.  With this choice of coordinates, then
in the contracting phase, the perturbation equation (\ref{perteq})
takes the form
\be \label{perteq2}
\Phi_k'' + \frac{1+2\nu}{\eta}\Phi_k' + c_s^2k^2\Phi_k \, = \, 0~,
\ee
where
\be \label{nuvalue}
\nu \, \equiv \, \frac{5+3w}{2(1+3w)}~.
\ee

Since (\ref{perteq2}) is a second order differential equation, there are
two linearly independent solutions. On super-Hubble scales, one mode
is constant (the ``D-mode"), the other is growing in a contracting phase (and
decreasing in an expanding phase). We call the second mode the
``S-mode". The general solution on super-Hubble scales is a linear
combination of the two modes and is hence given by
\be \label{asolc}
\Phi^{-}_k \, = \, D_{-}+\frac{ S_{-}}{(\eta)^{2\nu}}~,
\ee
where $D_{-}$ and $S_{-}$ are the mode coefficients.
If the equation of state is the same in the expanding phase after the
non-singular bounce, then the solution in the expanding period can
be written as
\be \label{asolc}
\Phi^{+}_k \, = \, D_{+}+\frac{ S_+}{(\eta)^{2\nu}}
\ee
with new mode coefficients $D_{+}$ and $S_{+}$.
In a non-singular bounce, the fluctuations can be smoothly evolved through
the bounce. Thus, the mixing matrix which relates the mode coefficients
in the expanding phase with those in the contracting phase can be
calculated \footnote{This calculation is rigorous in linear perturbation theory.
However, its physical validity is only assured if the fluctuations remain in the
linear regime throughout the evolution.}

The key question which arises when studying the transfer of fluctuations through
a non-singular bounce is whether the spectrum of the dominant mode in
the contracting phase, the $S_{-}$ mode, couples to the $D_{+}$ mode, the
dominant mode in the expanding phase. This issue was initially studied
by replacing the bounce region with a matching surface across which
the perturbations are connected making use of the Hwang-Vishniac \cite{HV}
(Deruelle-Mukhanov \cite{DM}) matching conditions. It was found
\cite{Lyth,FB1,KOST2} that the coupling is suppressed by a factor of $k^2$
on large wavelengths, i.e.
\be \label{coupling}
D_{+} \, = \, {\cal O}(1) D_{-} + {\cal O}(1) \bigl( \frac{k}{k^{*}} \bigr)^2 S_{-}
\ee
where $k^{*}$ is a normalization scale which is set by the microphysics,
i.e. is in the ultraviolet range. However, as pointed out in \cite{Durrer},
it is not valid to apply the matching conditions to fluctuations at an
interface between a contracting and an expanding phase because the
background does not satisfy the matching conditions. If the bounce
is non-singular, however, one does not need to make use of matching
conditions: the perturbations can be evolved continuously from the
contracting to the expanding phase. This was done in the context
of a non-singular regularization of Pre-Big-Bang cosmology in
\cite{Cartier}, and in the context of a non-singular Ekpyrotic model
in \cite{Tsujikawa2}. More recently, such calculations were carried
out for a non-singular mirage cosmology \cite{Omid}, in the non-singular
higher derivative bounce of \cite{Biswas} in \cite{Biswas3}, for a
quintom bounce model in \cite{Taotao1}, and more specifically in
the Lee-Wick bounce in \cite{LW}.

{F}or a non-singular bounce, one
can follow the fluctuations both numerically and analytically. To
obtain a good analytical approximation, one divides the
background time into three intervals - the contracting phase,
the bounce phase where the Hubble expansion rate can be
modelled as $H(t) = \alpha t$ (where $\alpha$ is a constant whose
value is set by the new physics which determines the bounce),
and the post-bounce
expanding phase. The duration of the bounce phase is
set by the scale determining the new physics which regulates
the bounce. The result of the works quoted above is that on
length scales larger than the duration of the bounce, the mode
mixing occurs as given by (\ref{coupling}). Thus, the contribution
of the dominant mode in the contracting phase is suppressed in
the dominant expanding phase mode function by a factor of $k^2$.
On the other hand, on scales short compared the bounce time,
there is no suppression of the coupling.

In the following, we will, without much loss of generality, assume that
the bounce is short from the point of view of cosmological scales
of interest. Thus, we will use the coupling given by (\ref{coupling}).
In \cite{LW}, we studied the evolution of perturbations which start
out as quantum vacuum fluctuations and discovered that it
is precisely quantum vacuum fluctuations which exit the Hubble radius
is a matter-dominated contracting phase which are scale-invariant
after the bounce. However, in the framework of a cosmological
model which starts out large and cold, there is no particular reason to
focus on quantum vacuum initial fluctuations. It is rather reasonable
to consider initial thermal perturbations. In this paper we will study
under which conditions on the background cosmology one obtains
a scale-invariant spectrum at later times starting from thermal
initial conditions.

Before starting the analysis, we remind the reader of a useful
formula for the time of Hubble radius crossing. Given an
equation of state parameter $w$, the scale factor evolves as
\be
a(t) \, \sim \, t^p~~,~~~{\rm with}~~p \, = \, \frac{2}{3(1+w)}~,
\ee
and yields the following relation for the conformal  time $\eta$
\be
\eta \, \sim \, t^{1-p}~.
\ee
The condition for Hubble radius crossing is
\be
k \, \sim \, a H
\ee
for a perturbation mode with fixed comoving
wavenumber $k$. This yields
\be
\eta_H(k) \, \sim \, k^{-1}~,
\ee
and thus
\be \label{scaling}
t_H(k) \, \sim \, k^{-\frac{1}{1-p}}~,
\ee
where the subscript ``$H$" denotes the moment of Hubble radius crossing.

\section{Thermal Fluctuations}

The method of calculating the spectrum of cosmological
perturbations at late times is the following. First, we compute
the matter fluctuations on sub-Hubble scales in the
contracting phase. Next, for any scale $k$, we compute
the induced metric fluctuations at the time $t_H(k)$ when the
scale exits the Hubble radius during the contracting phase.
In the third step, the metric fluctuations are evolved on super-Hubble
scales making use of the evolution equations for perturbations
discussed in the previous section. This is the standard way
of following the generation and evolution of cosmological
perturbations, as applied to inflationary cosmology in
early works (see e.g. \cite{BST} and \cite{BK}) and to string
gas cosmology \cite{NBV,BNPV2,RHBSGrev}. The method
of calculation reflects the fact that metric fluctuations are
sub-dominant on sub-Hubble scales, but that on super-Hubble
scales the matter fluctuations freeze out and the evolution
of the perturbations is driven by the metric.

The key constraint equation which relates matter and metric
fluctuations is the time-time component of
the perturbed Einstein equation
\be
-3{\cal H} ({\cal H}\Phi +\Phi') + \nabla^2\Phi \, =
\, 4\pi G  a^2\delta\rho~,
\ee
where $\delta\rho$ is the fluctuation of the energy density.
Note that all the three terms on the
left-hand-side of the above equation are of the same order of
magnitude at the Hubble radius crossing time. Therefore, up to
a constant of order $O(1)$, the power spectrum of the
metric perturbations is given by,
\bea \label{pthermal}
P_{\Phi}(k) \, &\equiv& \, \frac{1}{12 \pi^2} k^3 |\Phi(k)|^2 \\
&=& \, \frac{1}{4M_p^4} k^3 \frac{<\delta\rho(k)^2>}{(H(t_H(k))^4}~, \nonumber
\eea
where (in our case) the pointed brackets denote ensemble averaging in thermal
equilibrium. Making use of the Hubble radius crossing condition
$H(t_H(k)) = a^{-1}(t_H(k)) k$ and replacing the power spectrum of the
Fourier space energy density correlation function by the position space
correlation function we obtain:
\be \label{pthermal2}
P_{\Phi}(k)(t_H(k)) \, = \, \frac{1}{4M_p^4} k^{-4} a^4(t_H(k)) <\delta\rho^2>|_{R(k)}~,
\ee
where $<\delta\rho^2>|_{R(k)}$ (to be evaluated at Hubble radius crossing)
is the position space energy density
fluctuation correlation function is a sphere of radius $R(k)$, where $R(k)$ is
the physical length corresponding to the co-moving momentum scale $k$.

In a system which is in thermal equilibrium, the correlation function of
the energy density is given by
\be \label{prho}
<\delta\rho^2>|_{R(k)} \, \equiv \, C_V(R) \frac{T^2}{R^6}~,
\ee
where $C_V(R)$ is the heat capacity in a sphere of radius $R$
and is defined in terms of the expectation value of the
internal energy as
\be \label{CR}
C_V(R) \, \equiv \, \frac{\partial}{\partial T}<E>~.
\ee

\subsection{Thermal Particle Fluctuations}

In this subsection we consider fluctuations in a gas of point particles
with an arbitrary equation-of-state $w_r$. In this case, from the
stress-energy conservation equation it follows that the
energy density and the temperature change as a function of the
scale factor as follows:
\bea
\rho_r \, &\sim& \, a^{-3(1+3w_r)} \\
T \, &\sim& \,  a^{-3w_r}~,
\eea
which yields
\be
\rho_r \, \sim \, T^{1+\frac{1}{w_r}} \, ,
\ee
and so we get the heat capacity
\be \label{CR1}
C_V(R) \, = \, R^3\frac{\partial\rho}{\partial T} \, \sim \, R^3T^{\frac{1}{w_r}}~.
\ee

Inserting (\ref{prho}) and (\ref{CR1}) into (\ref{pthermal}) and
applying $R\sim 1/H$, we obtain the power spectrum for metric
perturbations at the Hubble radius crossing time $t_H(k)$:
\be
P_{\Phi}(k) \, \sim \, T_H^{2+\frac{1}{w_r}}H_{t_H(k)}^{-1} \,\sim \,
k^{\frac{1-3p(1+2w_r)}{p-1}}~,
\ee
where we have used the relations
\bea
T_H(k) \ &\sim& \,  a^{-3w_r}\sim k^{\frac{3w_rp}{1-p}}~,~{\rm and} \\
H_{t_H(k)} \, &\sim& \, k^{\frac{1}{1-p}}~.
\eea

Since the power spectrum of the constant mode $P_D(k)$ is the same
as that of $\Phi$ at Hubble radius crossing, it scales as
\bea
P_D(k) \, &\sim& \, k^{n_D} \\
n_D \, &=& \, \frac{1-3p(1+2w_r)}{p-1}~. \nonumber
\eea
The power spectrum of the growing mode $P_S(k)$ is the spectrum
of $\Phi$ at Hubble radius crossing modulated by the factor
$\eta_H(k)^{4\nu}$. This yields
\bea
P_S(k) \, &\sim& \, P_D(k)k^{-4\nu} \, \sim \, k^{n_S} \\
n_S \, &=& \, \frac{3-p(1+6w_r)}{p-1}. \nonumber
\eea

There are two possibilities to obtain a scale-invariant
spectrum after the bounce. The first is if the D-mode in
the contracting phase is scale-invariant (recall that
the contribution of the contracting phase D-mode to the
expanding phase D-mode is not suppressed). The
second is that the S-mode in the contracting phase
has a spectrum proportional to $k^{-4}$. Since the
contribution of the contracting phase S-mode to the
expanding phase D-mode is suppressed by $k^4$ in
the power spectrum (see (\ref{coupling}), this then yields
scale-invariance of the dominant mode in the expanding
phase. Thus, the two possibilities to obtain a scale-invariant
spectrum are
\bea
\label{thermal1} n_D=0~,~~n_S+4>0~;\\
{\rm or}~~ \label{thermal2} n_D>0~,~~n_S+4=0~.
\eea

In order for the D-mode to provide a scale-invariant spectrum, we require
\be \label{wr1}
w_r \, = \, \frac{1}{4}(w - 1) \, .
\ee
This is a solution provided that the S-mode yields a blue spectrum which
requires
\be
\frac{w + 4 w_r - 1}{w + 1/3} \, \geq \, 0
\ee
which is satisfied if $w > 1$. Note that for such a background, the
anisotropic stress grows less fast than the background energy
density. Thus, this background is stable towards anisotropic
stress fluctuations. Note also that if we were to demand that
the equation of state of the background and of the fluctuations
is the same, we would obtain $w = - 1/3$ which is not a
physical solution since it is the borderline solution between an
acelerating and a decelerating background, and no scales
exit the Hubble radius during the contracting phase.

For the growing mode (S-mode) in the contracting phase to
provide a scale-invariant spectrum in the expanding phase,
the condition on the two equations of state is
\be \label{wr1}
w_r \, = \, \frac{1}{4}(w - 1)~,~~
\ee
and the dominance of the S-mode requires
\be
- \frac{1}{3} \, < \, w \, < \, 1~.
\ee

In the special case that the radiation is normal
radiation with $w_r=\frac{1}{3}$, the only possibility
to obtain a scale-invariant spectrum is via the D-mode
in the contracting phase. This requires
\be
w \, = \, \frac{7}{3} \, .
\ee

\subsection{Fluctuations in Holographic Cosmology}

Gibbons-Hawking radiation \cite{GH}, originally discovered in
studies of thermodynamics in de-Sitter space, has recently
been studied extensively in the context of
developments in string theory, in particular in light
of the role of holography in string theory \cite{FS}. The key point
is that quantities such as the entropy and the energy do not
scale extensively with the size of the volume, but increase
as the area \cite{FS}. Thus, the average energy is
\be
<E> \, = \, T R^2M_p^2~,
\ee
which gives a special heat capacity
\be \label{CR2}
C_V(R) \, \sim \, R^2 M_p^2~.
\ee

If we combine Eqs. (\ref{pthermal}), (\ref{prho}) and
(\ref{CR2}), and use the definition of the Gibbons-Hawking temperature
associated with the instantaneous Hubble radius
\be
T \, = \, \frac{1}{R} \, ,
\ee
where $R$ is taken to be the Hubble radius $R\sim1/H$, then
the power spectrum for $\Phi$ at the Hubble
radius crossing is given by
\bea \label{hol}
P_{\Phi}(k, t_H(k)) \, &\sim& \, (\frac{T_H(k)}{M_p})^2 \\
&\sim& \, (\frac{H(t_H(k))}{M_p})^2 \, \sim \, k^{2/(1 - p)} \, .\nonumber
\eea
Thus, a scale-invariant spectrum from the D-mode can be achieved
in the limit $p \rightarrow \infty$, which corresponds to an inflationary
contraction. However, in this case scales are not exiting the Hubble
radius during the contracting phase, and thus it does not make sense
to consider thermal fluctuations in this context since thermal equilibrium
cannot be established on super-Hubble scales.

The contribution of the S-mode scales as
\be
P_{\Phi}(k, t) \, \sim \, k^4 \eta_H(k)^{4 \nu} k^{\frac{2p}{1 - p} + 2} \, .
\ee
Inserting the relation (\ref{nuvalue}) for $\nu$ in terms of $p$
we find that the condition for scale-invariance is
\be
w \, = \, 0 \, .
\ee

As a consistency check, we note that, as follows from (\ref{hol}),
for $p = 2/3$ the D-mode has a red spectrum. Hence, this
is indeed a background equation of state for which thermal
fluctuations of a holographic gas yields a scale-invariant spectrum.

\subsection{Thermal Fluctuations of a String Gas}

For completeness, we will also add an analysis for thermal
string gas fluctuations. If space is compact, then the specific heat
capacity of a gas of closed strings also satisfies the holographic
scaling
\be
C_V(R) \, \sim \, R^2 M_{pl}^2\, .
\ee
Inserting this relation into (\ref{pthermal2}), (\ref{prho}) and (\ref{CR})
we find
\be
P_{\Phi}(k, t_H(k)) \, \sim \, a^4(t_H(k)) T^2(t_H(k)) \, .
\ee

In the case of string gas cosmology \cite{BV,NBV,BNPV2,RHBSGrev},
fluctuations exit the Hubble radius at the end of a quasi-static Hagedorn
phase. Thus, both $a(t_H(k))$ and $T(t_H(k))$ are almost independent
of $k$, and a scale-invariant spectrum (with a slight red-tilt) results.

If we forget about the Hagedorn background (which cannot be
described in terms of the Einstein or dilaton gravity background
equations), and simply couple a string gas to the non-singular
background geometry discussed in this paper, and use
$T(t) \sim a(t)^{-1}$, then the D-mode of $\Phi$ yields
a scale-invariant spectrum if $p = 0$. This is consistent with
the Hagedorn phase of string gas cosmology. We also
find that the S-mode yields a scale-invariant contribution to
the post-bounce spectrum of $\Phi$ if $p = 1/4$, but for this
value of $p$, the contribution from the D-mode has a red tilt
and hence dominates.

\section{Non-Gaussianities from Thermal Fluctuations}

Due to the non-linearities in the theory, the fluctuations are
not perfectly Gaussian. A lot of recent interest has focused
on calculating the non-Gaussianities as manifested in the
three-point function (see e.g. \cite{NGrev} for a review). In
single field slow-roll inflation models the amplitude of the
predicted non-Gaussianities is suppressed by the slow-roll
parameter. In models with a contracting phase, however,
the induced non-Gaussianities are typically much larger
(see e.g. \cite{EkpNG} for studies in the context of
the Ekpyrotic scenario).

As studied recently in \cite{Cai2009},
the non-Gaussianities as measured by the three-point function
are of order one in the non-singular matter bounce scenario in
which the fluctuations are of quantum vacuum origin. The
key facts that lead to this result are firstly that there is no
suppression of the non-Gaussianities by slow-roll parameters,
and secondly that fluctuations grow on super-Hubble scales
in the contracting phase, which results in a larger time interval
determining the amplitude of the effects and in different terms
dominating the shape of the three point function.

In this section we calculate the non-Gaussianity estimator $f_{NL}$
in bouncing cosmologies with thermal fluctuations. We will
use the formalism to compute non-Gaussianities of thermal
origin which has been developed in \cite{Xue1} and which was
applied to estimate the non-Gaussianities in string gas cosmology
in \cite{Xue2} and to holographic cosmology in \cite{Ling}.
We will calculate the following non-Gaussianity estimator for
fluctuations on a scale $k$:
\be \label{fthermal}
f_{NL}(k) \, \equiv \, \frac{5}{18\sqrt{k^3}}\frac{<\zeta_k^3>}{<\zeta_k^2>^2}~,
\ee
where $\zeta$ is the curvature fluctuation in co-moving gauge.
Thus if we obtain the two-point and three-point correlators of
metric perturbations originating from thermal fluctuations, the
non-Gaussianity parameter can be calculated \footnote{Implicitly, we
are estimating the non-Gaussianities in the equilateral limit. In general,
the three point function is a function of three momenta whose vector sum
is zero. Here, we are taking the magnitudes of all momenta to be
comparable, namely $k$.}.

The expression for the two-point correlation function was given in
(\ref{prho}). The three-point correlation function of
thermal fluctuations in an equilibrium ensemble is
\be \label{drho3}
<\delta\rho^3> \, \equiv \, -\frac{1}{R^9}\frac{\partial^3\ln{Z}}{\partial\beta^3}
\, = \, \frac{T^2}{R^9}\frac{\partial}{\partial{T}}(C_RT^2)~,
\ee
where $Z$ is the partition function and $\beta$ is the inverse temperature.

\subsection{Normal Radiation}

For normal radiation with equation of state $w_r=1/3$,
we need the background equation of state of the universe in
the contracting phase to be $w = 7/3$ in order to obtain a
scale-invariant spectrum of fluctuations, as studied in Section 3.
In this case, the heat capacity can be expressed as
\be
C_V(R) \, = \, c_v R^3T^3 \, ,
\ee
where $c_v$ is determined by the background initial
conditions and here is
treated as a constant of order $O(1)$. From this,
we obtain the following expressions for the two-point and
three-point correlation functions of the density perturbations
\begin{eqnarray}
<\delta\rho^2> \, &=& \, \frac{c_vT^5}{R^3}~,\\
<\delta\rho^3> \, &=& \, \frac{5c_vT^6}{R^6}~.
\end{eqnarray}

From Eq. (\ref{pthermal}) we find the relation
\be \label{Phit}
\Phi \, = \, \frac{\delta\rho}{2M_p^2H^2}~,
\ee
which is valid for any thermal system. In Section 3 we have
seen that for the background considered here, the dominant mode
after the bounce is seeded by the constant mode (the D-mode)
in the contracting phase. Hence, the relation between $\Phi$
and $\zeta$ is given by
\be
\zeta \, = \, \frac{5+3w}{3+3w}\Phi_D \, = \, \frac{6}{5}\Phi_D~.
\ee

Combining the above equations, and going from Fourier
space to position space via
\be
\delta \rho \, =   2^{-1/2} \pi^{-1}  \delta \rho(k) \, ,
\ee
where the left-hand side represents the root mean square density
fluctuation corresponding to the co-moving scale $k$,
finally the non-Gaussianianity estimator takes the form
\be \label{fnl0}
f_{NL} \, = \,
\frac{25}{54\sqrt{2}\pi}M_p^2H_{t_H(k)}^2\frac{<\delta\rho^3>}{<\delta\rho^2>^2}\bigg|_{t_H(k)}~,
\ee
where the right hand side is evaluated at the Hubble radius crossing time.
Inserting the above expressions for the density two and three point functions we
get
\be \label{fnl}
f_{NL} \, = \,
\frac{125}{54\sqrt{2}\pi} \frac{M_p^2H_{t_H(k)}^2}{c_v T^4(t_H(k))} \, .
\ee
Noticing that for $w_r = 1/3$ we have $T\propto a^{-1}$,
inserting $H(t_H(k)) \sim t^{-1}(t_H(k))$ and making use of (\ref{scaling}) we obtain
\be
f_{NL} \, \sim \, k^{2 \frac{1 - 2p}{1 - p}} \, ,
\ee
and the coefficient is such that for scales exiting the Hubble radius right
before the bounce (when all quantities on the right hand side of (\ref{fnl})
are of the order of the Planck mass - assuming that the energy scale
at the bounce is given by the Planck mass) the amplitude of $f_{NL}$ is
of order $1$. Inserting the value $p = 1/5$ obtained for $w = 7/3$ we
finally obtain
\be
f_{NL}(k) \, \sim \, {\cal{O}}(1) (\frac{k}{k_B})^{3/2} \, ,
\ee
where $k_B$ is the value of $k$ for which the wavelength exits the
Hubble radius immediately before the bounce.

The above analysis shows that the non-Gaussianities in a bouncing
cosmology in which the fluctuations are seeded by particle thermodynamic
perturbations are Poisson suppressed on wavelengths larger than
those exiting the Hubble radius immediately before the bounce.
Thus, the non-Gaussianities are very different from what is obtained
\cite{Cai2009} in the case of a matter bounce with quantum vacuum
initial perturbations. Besides the Poisson suppression of thermal
non-Gaussianities (by the central limit theorem, on large scales the
thermal fluctuations must approach a Gaussian), an important reason
for the difference is that in the case of the ``matter bounce" of \cite{Cai2009}
the S-mode is responsible for the final fluctuations. The S-mode grows
on scales larger than the Hubble radius in the contracting phase, thus
leading to an enhancement of the non-Gaussianities.

\subsection{An Estimate of $f_{NL}$ for Gibbons-Hawking Radiation}

In the case of a holographic gas we expect the resulting non-Gaussianities
to be much larger since for each scale the thermal correlation length
is taken to be equal to the Hubble radius at the time that the scale exits
the Hubble radius. Hence, we do not expect a Poisson suppression factor,
and we expect non-Gaussianities to be of order $1$. This expectation
is verified by an explicit computation.

The starting point is the expression $C_V(R) = c_vM_p^2R^2$ for the
heat capacity, which leads to the following expression for the
two-point and three-point correlation functions of energy density
perturbations
\bea \label{drhot2}
<\delta\rho^2> \, &=& \, \frac{c_vM_p^2T^2}{R^4} \, ,\\
<\delta\rho^3> \, &=& \, \frac{2c_vM_p^2T^3}{R^7}~.
\eea
Inserting these expressions into (\ref{fnl0}) we obtain
\be
f_{NL} \, = \, \frac{25}{54 \sqrt{2} \pi} \frac{1}{c_v} \frac{R H^2}{T} \, ,
\ee
where all quantities on the right hand side are to be evaluated
at Hubble radius crossing.

Making use of the Gibbons-Hawking relation $T\simeq1/R\simeq
H$, we then have
\be
f_{NL} \, \simeq \, \frac{1}{c_v}~,
\ee
which shows the non-Gaussianity is of order $O(1)$
and scale independent.

Note that the sign of $f_{NL}$ is positive in the case of thermal fluctuations
considered here. This is different from the negative sign obtained in
the matter bounce scenario with vacuum initial conditions \cite{Cai2009}.

\section{Conclusions}

In this paper we have studied the possibility of obtaining a scale-invariant
spectrum of fluctuations from thermal initial conditions in the
context of a non-singular bouncing cosmology. We classified the conditions on the
equation of state $w_r$ of the thermal radiation and the equation of
state $w$ of the background which yield a scale-invariant spectrum.

In the case of regular particle radiation with $w_r = 1/3$ we find the
condition $w = 7/3$ for the background. Such as equation of state
can be realized by supposing that the background is determined
by a scalar field with a negative exponential potential, similar to
what is assumed in Ekpyrotic cosmology \cite{Ekp}. In this case, the
fluctuations in radiation would be entropy fluctuations which would
then seed adiabatic fluctuations with the same spectral index (note
that the primordial adiabatic fluctuations induced by the scalar
field are blue and hence will not dominate for long wavelength
modes \cite{Lyth,FB1}).

In the case of a Gibbons-Hawking radiation we find that an
equation of state $w = 0$  (a matter bounce) yields a
scale-invariant spectrum of cosmological perturbations.

Finally, to obtain a scale-invariant spectrum of cosmological
perturbations with a  string gas requires a quasi-static
early phase. This matches with what is usually assumed
in string gas cosmology \cite{RHBSGrev}.

We have also considered the non-Gaussianities of thermal fluctuations.
In the case of thermal particle fluctuations we find that
the non-Gaussianities are Poisson-suppressed on large
scales. They are of the order $1$ only on microscopic
length scales, scales for which the thermal correlation length
at the time of Hubble radius crossing is comparable with the
Hubble radius itself. However, in the case of
radiation satisfying the Gibbons-Hawking distribution,
the non-Gaussianity is of order $1/c_v$ (where $c_v$ is the
constant which determines the specific heat capacity)
and approximately scale independent.

\section*{Acknowledgments}

We would like to thank Yi Wang for discussions. R.B. wishes to
thank the Theory Division of the Institute of High Energy Physics
(IHEP) and the KITPC for their hospitality and financial support.
R.B. and W.X. are also supported by an NSERC Discovery Grant and
by the Canada Research Chairs Program. The research of Y.C. and
X.Z. is supported in part by the National Science Foundation of
China under Grants No. 10533010 and 10675136, and by the Chinese
Academy of Sciences under Grant No. KJCX3-SYW-N2

\end{document}